\documentclass[1p]{elsarticle}
\usepackage{hyperref}
\usepackage[utf8]{inputenc}
\usepackage{amsfonts}
\usepackage{amssymb}
\usepackage{mathtools}

\journal{Applied Mathematics Letters}

\begin{document}

\numberwithin{equation}{section}

\newcommand{\inv}[1]{\ensuremath{\frac{1}{#1}}}         
\newcommand{\tinv}[1]{\ensuremath{\tfrac{1}{#1}}}       
\newcommand{\order}[1]{\ensuremath{\mathcal{O}\left(#1\right)}}        
\newcommand{\me}[1]{\ensuremath{\mathrm{e}^{#1}}}       
\newcommand{\Ad}[1]{{\mathrm{Ad}} #1}                     
\newcommand{\eg}{e.g.,\ }
\newcommand{\Eg}{E.g.,\ }
\newcommand{\ie}{i.e.,\ }

\begin{frontmatter}
  
\title{Lie symmetry analysis and exact solutions of the one-dimensional heat equation
with power law diffusivity}

\author[myaddress]{Tobias F. Illenseer}
\ead{tillense@astrophysik.uni-kiel.de}
\address[myaddress]{Institut f\"ur Theoretische Physik und Astrophysik, %
  Christian-Albrechts-Universit\"at zu Kiel, Leibnizstr. 15, 24\,118 Kiel, Germany}

\begin{abstract}
  A heat equation with non-constant diffusivity depending as a power law on the
  spatial variable is analysed using Lie's method to identify classical point
  symmetries. It is shown that the group invariant solutions of a
  four-dimensional symmetry subgroup can be decomposed into three different
  classes. These admit explicit solutions which can either be expressed in
  terms of Bessel functions, confluent hypergeometric functions or Coulomb wave
  functions.
\end{abstract}

\begin{keyword}
  Heat equation with power law diffusivity \sep Lie-group methods \sep exact solutions
  \MSC[2010] 35K05 \sep 76M60 \sep 85A30
\end{keyword}

\end{frontmatter}


\section{Introduction}

The purpose of this paper is to find Lie point symmetries and associated exact
solutions of the one-dimensional heat equation with power law diffusivity
\begin{equation}
  \label{eqn:pde}
  u_t - x^{2-1/a}\, u_{xx} = 0,\quad a \in \mathbb{R}\backslash\{0\}.
\end{equation}
If $a=1/2$ one obtains the heat equation with constant diffusivity whose Lie
point symmetries are well-known \cite{bluman1974,olver1986}. For $a\to\pm\infty$
the equation can be transformed to the constant coefficient heat equation
\cite{polyanin2002}. The transformation $y = x^{-1}, v=y\, u(t,x(y))$ yields an
equation of similar form for $v(t,y)$ with the parameter $a$ replaced by $-a$
\cite{polyanin2002}. Hence in the following we assume that
$a\in\mathbb{R}^+\backslash\{1/2\}$.

A second order partial differential equation (PDE) of the above type is used
in stationary two-dimensional diffusion boundary layer problems to model the
evaporation of particles into a turbulent medium \cite{polyanin2002,sutton1943}.
It also plays a major role in the theory of accretion discs, an astrophysical
research field which deals with rotating fluid flows of gaseous discs under
the influence of friction and gravity \cite{lyndenbell1974}.

The general solution of equation (\ref{eqn:pde}) in terms of Green's function
is well known. It is obtained with help of the transformation
$y=2ax^{1/(2a)}, v(t,y)=x^{-a}\,u(t,x(y))$ for $a\in\mathbb{R}^+$
and application of Hankel transforms of order $a$ with respect to the new
variable $y$ \cite{polyanin2002,sutton1943,lyndenbell1974}. Some exact solutions
including similarity solutions are also known \cite{polyanin2002,lyndenbell1974,rafikov2016}.

\section{Generator of the Lie group of point transformations}
\label{app:lie_group_generator}

Let's consider the one-parameter Lie group of infinitesimal point transformations
\begin{equation}
  \label{eqn:lie_group}
  \hat{t} = t + \varepsilon\tau(t,x,u) + \order{\varepsilon^2},~
  \hat{x} = x + \varepsilon\xi(t,x,u) + \order{\varepsilon^2},~
  \hat{u} = u + \varepsilon\eta(t,x,u) + \order{\varepsilon^2}
\end{equation}
and its generator
\begin{equation}
  \label{eqn:generator_general}
  X=\tau\partial_t+\xi\partial_x+\eta\partial_u.
\end{equation}
As usual one computes the prolongation up to second order in the derivatives of
$u$
and applies this operator to the PDE (\ref{eqn:pde}) to derive the linearized
symmetry condition \cite{bluman1974,olver1986,hydon2000}
\begin{equation}
  \label{eqn:symmetry_condition}
  \eta^{(t)} - (2-1/a)x^{1-1/a} \xi  u_{xx} - x^{2-1/a} \eta^{(xx)} = 0
\end{equation}
with
\begin{align}
  \eta^{(t)} & = \eta_t - \xi_t u_x + (\eta_u - \tau_t) u_t - \xi_u u_x u_t - \tau_u u_t^2 \\
  \eta^{(xx)} & = \eta_{xx} + (2\eta_{xu} - \xi_{xx}) u_x - \tau_{xx} u_t
    + (\eta_{uu} - 2\xi_{xu}) u_x^2 - 2\tau_{xu} u_x u_t - \xi_{uu} u_x^3 \nonumber \\
    &\quad- \tau_{uu} u_x^2 u_t + (\eta_u-2\xi_x)u_{xx} - 2\tau_xu_{xt} - 3\xi_u u_x u_{xx}
    - \tau_u u_t u_{xx} - 2\tau_u u_x u_{xt}.
\end{align}
Inserting these expressions in (\ref{eqn:symmetry_condition}),
collecting terms of like derivatives of $u$ and equating the coefficient
functions with zero yields the determining equations
\begin{align}
  \label{eqn:determining_equations1}
  \tau_u = 0,\quad \tau_x = 0,\quad \xi_u = 0,\quad \eta_{uu} = 0,\quad
  \eta_t - x^{2-1/a}\eta_{xx} = 0, \\
  \label{eqn:determining_equations2}
  \xi_t - x^{2-1/a}\xi_{xx} = -2 x^{2-1/a}\eta_{xu},\quad
  2x\xi_x - (2-1/a)\xi = x\tau_t.
\end{align}
This is already a reduced set of equations in which some obvious simplifications
were applied to eliminate redundancies. From the first four equations in
(\ref{eqn:determining_equations1}) one concludes immediately that
$\tau=\tau(t)$, $\xi=\xi(t,x)$, and $\eta=V(t,x)u+W(t,x)$.
Inserting the last expression for $\eta$ in the fifth equation shows that
$V(t,x)$ and $W(t,x)$ must be particular solutions of the original PDE. If
$a\ne\pm 1/2$ the system is solved by
\begin{equation}
  \label{eqn:point_symmetry_transformations}
  \begin{split}
  \tau(t) &= k_1 + k_2 t + k_3 t^2,\quad
  \xi(t,x) = a\bigl(k_2 + 2 k_3 t\bigr)x,\\
  \eta(t,x,u) &= \Bigl(k_4 - k_3 \bigl((1-a)t+a^2 x^{1/a}\bigr)
    \Bigr)u + W(t,x)
  \end{split}
\end{equation}
where the $k_j$ are arbitrary constants. The Lie algebra associated with this
symmetry group of infinitesimal point transformations is spanned by
\begin{equation}
  \label{eqn:generators}
  \begin{split}
  \Bigl\{X_W &= W(t,x)\partial_u : W_t = x^{2-1/a}W_{xx}\Bigr\},\quad
  X_1 = \partial_t,\quad X_2 = t\partial_t + a x \partial_x, \\
  X_3 &= t^2\partial_t + 2 a t x \partial_x - \bigl((1-a)t+a^2 x^{1/a}\bigr) u \partial_u,\quad
  X_4 = u\partial_u.
  \end{split}
\end{equation}
In the subsequent sections we will only consider symmetries of the 
four-dimen\-sional subalgebra $\mathcal{L}_4$ generated by $\{X_1,X_2,X_3,X_4\}$
whose non-vanishing structure constants $c^k_{ij}$ which are defined according to
\begin{equation}
  [X_i,X_j] = \sum_{k=1}^4 c^k_{ij} X_k\quad\textnormal{with}\quad i,j=1,\dots,4
\end{equation}
are given by
\begin{equation}
  \label{eqn:commutators}
  [X_1,X_2] = X_1,\quad [X_1,X_3] = 2 X_2 + (a-1) X_4,\quad
  [X_2,X_3] = X_3.
\end{equation}
\textbf{Remark.} The subalgebra spanned by $\{X_1,X_2,X_3\}$ is the special linear
Lie algebra $\mathfrak{sl}(2)$. If $a=1$ the algebra $\mathcal{L}_4$ decomposes
into the $\mathfrak{sl}(2)$ and the one-dimensional subalgebra generated by $X_4$.

\section{Optimal system of generators}
\label{sec:optimal_system}

Before proceeding with the construction of the group invariant solutions we
first examine the structure of the algebra and deduce an optimal system of
generators. This allows for classification of the solutions into classes of
equivalent solutions. Within each class we try to find the most simple
representative to which any other solution in the class is related by
the adjoint group action given by the Lie series \cite{olver1986,ovsiannikov1982}
\begin{equation}
  \label{eqn:adjoint_representation}
  \widetilde{X} = \Ad{(\exp(\varepsilon X_i))} X = X - \varepsilon [X_i,X]
    + \frac{\varepsilon^2}{2} [X_i,[X_i,X]] - \cdots.
\end{equation}
This maps the generator
\begin{equation}
  \label{eqn:generator_4dim}
  X = \sum_{j=1}^4 k_j X_j
\end{equation}
having components $k_j$ with respect to the basis $\{X_1,X_2,X_3,X_4\}$ to some
other generator $\widetilde{X}$ with components $\widetilde{k}_j$. To obtain the
optimal representative one tries to make as many of the $\widetilde{k}_j$ as
possible zero. Usually it is necessary to apply different adjoint maps successively.
Further simplification can be achieved by rescaling $X$.

The crucial point is that this mapping does not necessarily connect any two
generators. Instead there are restrictions due to the existence of invariants
of the adjoint action, \ie vector-valued functions $\Phi(X)$ satisfying 
the following system of linear first order PDEs:
\cite{hydon2000,beltrametti1966}
\begin{equation}
  \label{eqn:invariants_adjoint_action_pde}
  \sum_{j,m=1}^4 k_m c_{mi}^j \partial_{k_j} \Phi = 0
    \quad\textnormal{for}\quad i=1,\dots,4.
\end{equation}
The solutions of the system (\ref{eqn:invariants_adjoint_action_pde})
with $c_{mi}^j$ from (\ref{eqn:commutators}) can be obtained using the method
of characteristics which gives two invariants in this case
\begin{equation}
  \label{eqn:invariants_adjoint_action}
  \Phi_1 = k_2^2 - 4 k_1 k_3,\quad 
  \Phi_2 = \begin{cases}
             k_4 &\textnormal{if}\quad a=1 \\
             k_2 + \tfrac{2}{1-a} k_4 &\textnormal{if}\quad a\in\mathbb{R}^+\backslash\{\tinv{2},1\}
           \end{cases}
\end{equation}
The existence of these invariants is the actual reason for the
group invariant solutions beeing subdivided into subsets of equivalent
solutions.

Although these invariants are preserved under the adjoint group action they are
affected by rescaling $X$. If, \eg $\Phi_1=0$ this has no effect, but if $\Phi_1\ne0$
it is rescaled by a positive constant. In case of $\Phi_2$ one can also alter
the sign. Thus we must distinguish the six cases $\Phi_1>0$, $\Phi_1<0$, $\Phi_1=0$
combined with either $\Phi_2\ne 0$ or $\Phi_2=0$ to obtain the optimal system
of generators
\begin{equation}
  \label{eqn:optimal_system}
  X_1+\mu X_4,\quad X_2+\mu X_4,\quad X_1+X_3+\mu X_4,\quad X_4
\end{equation}
with $\mu\in\mathbb{R}$. The derivation for $a=1$ can be found in \cite{hydon2000},
examples 10.2 and 10.4. The considerations for $a\ne 1$ are similar but need a
little more case-by-case analysis.

\section{Differential invariants and group invariant solutions}
\label{sec:group_invariant_solutions}

A solution $u(t,x)$ of the PDE (\ref{eqn:pde}) defined implicitly by $F(t,x,u)=0$
is invariant with respect to the four-dimensional Lie group of point transformations
(\ref{eqn:point_symmetry_transformations}) with $W(t,x)=0$ if it satisfies the
invariant surface condition $X F = 0$ with $X$ given by (\ref{eqn:generator_4dim}).
Accordingly we derive the group invariant solutions admitted by the generators
(\ref{eqn:optimal_system}) for each subgroup of the optimal system. Since
the invariant surface condition yields a linear first order PDE its solution
is obtained solving the characteristic equations which gives the differential
invariants.

The group invariant solution with respect to the generator $X_4=u\partial_u$ 
is only the trivial solution $u=0$. Thus three different classes of non-trivial
group invariant solutions remain.

\textit{Case 1.} Generator $X=X_1+\mu X_4$\\
This group comprises the stationary and the separable solutions \cite{polyanin2002}.
The differential invariants are $x$ and $u\me{\mu t}$. If $\mu=0$ the
group invariant solutions are the stationary solutions $u=C_1 x + C_2$.
Otherwise we define $\mu=\pm\kappa^2$ with $\kappa> 0$ and obtain the separable
solutions
\begin{equation}
  \label{eqn:group_invariant_solution_case1}
  u(t,x) = \me{\pm\kappa^2 t} \xi^a y(\xi),\quad\textnormal{with}\quad \xi(x) = 2\kappa a x^{1/2a}
\end{equation}
where $y(\xi)$ is a solution of the (modified) Bessel differential equation
\cite{abramowitz1964}
\begin{equation}
  \label{eqn:bessel_differential_equation}
  \xi^2 y'' + \xi y' + \bigl(\mp\xi^2-a^2\bigr) y = 0.
\end{equation}

\textit{Case 2.} Generator $X=X_2+\mu X_4$\\
This yields the scale invariant similarity solutions recently found by
\cite{rafikov2016}. Some particular cases are also listed in \cite{polyanin2002,lyndenbell1974}.
The differential invariants are $x/t^a$ and $u/t^\mu$ and the group invariant
solutions are given by
\begin{equation}
  \label{eqn:group_invariant_solution_case2}
  u(t,x) = t^\mu \xi^a \me{-\xi} y(\xi),\quad\textnormal{with}\quad \xi = a^2 x^{1/a} t^{-1}
\end{equation}
where $y(\xi)$ is a solution of the confluent hypergeometric differential equation
\cite{abramowitz1964}
\begin{equation}
  \label{eqn:confhyp_differential_equation}
  \xi y'' + \bigl(1+a-\xi\bigr) y' - (1+\mu)\, y = 0.
\end{equation}

\textit{Case 3.} Generator $X=X_1 + X_3 +\mu X_4$\\
With the differential invariants
$x/(1+t^2)^a$ and
\begin{equation}
  u\,\left(1+t^2\right)^\frac{1-a}{2}\exp{\left(-\mu\arctan{t} + a^2 \frac{x^{1/a}\,t}{1+t^2}\right)}.
\end{equation}
the group invariant solutions of this class become
\begin{equation}
  \label{eqn:group_invariant_solution_case3}
  u(t,x) = \bigl(a\,x^{1/2a}\bigr)^{a-1} \me{\left(\mu\arctan{t} - \xi t\right)}\,
    \,y(\xi),\quad\textnormal{with}\quad \xi = a^2 x^{1/a} (1+t^2)^{-1}
\end{equation}
where $y(\xi)$ is a solution of the Coulomb wave equation \cite{abramowitz1964}
\begin{equation}
  \label{eqn:coulomb_wave_equation}
  y'' + \biggl(1 - \frac{\mu}{\xi} - \frac{\ell(\ell+1)}{\xi^2}\biggr) y = 0,
  \quad\textnormal{with}\quad \ell = \frac{a}{2}-\inv{2}.
\end{equation}
More general three-parameter solutions which are invariant with respect
to the full group action can be obtained from these solutions applying suitable
group transformations \cite{olver1986,hydon2000}. If $F(t,x,u)=0$ is one of the 
basic solutions given above then $F(\hat{t},\hat{x},\hat{u})=0$ with
$\hat{t}=\me{\varepsilon X_j} t,\ \hat{x}=\me{\varepsilon X_j} x,\ \hat{u}=\me{\varepsilon X_j} u $
is also a group invariant solution. One easily proofs with help of the Lie
series, that
\begin{align}
  \label{eqn:finite_transformations_1}
  \me{\varepsilon X_1} t &= t + \varepsilon, &
  \me{\varepsilon X_1} x &= x, &
  \me{\varepsilon X_1} u &= u \\
  \label{eqn:finite_transformations_2}
  \me{\varepsilon X_2} t &= \me{\varepsilon} t, &
  \me{\varepsilon X_2} x &= \me{\varepsilon a} x, &
  \me{\varepsilon X_2} u &= u, \\
  \label{eqn:finite_transformations_3}
  \me{\varepsilon X_3} t &= t \sum_{j=0}^\infty (\varepsilon t)^j, &
  \me{\varepsilon X_3} x &= \me{\varepsilon 2 a t} x, &
  \me{\varepsilon X_3} u &= \me{-\varepsilon\bigl((1-a)t+a^2 x^{1/a}\bigr)} u.
\end{align}
Since $X_4$ commutes with all generators, this transformation (which just
rescales $u$) does not yield any new solutions. The generator $X_3$ induces
a transformation of time $t$ which only maps to finite values if $|\varepsilon t| < 1$.
Otherwise the geometric series in (\ref{eqn:finite_transformations_3}) diverges.
Thus depending on the time-dependence, solutions obtained applying this
transformation may become infinitely large or vanish everywhere if $t$ exceeds a
certain value.

It can be shown that time shifts and scalings applied to separable solutions
map to other separable solutions. Therefore only transformation
(\ref{eqn:finite_transformations_3}) generates new solutions in this case.
Scale invariant similarity solutions can be transformed to new solutions
applying time shifts (\ref{eqn:finite_transformations_1}) and the transformations
(\ref{eqn:finite_transformations_3}). The third class of basic solutions 
admits new solutions if scalings (\ref{eqn:finite_transformations_2}) or
transformations generated by $\varepsilon X_1 + X_3$ with $\varepsilon\ne 1$ are
applied.

\section{Conclusions}
\label{sec:conclusions}

The heat equation with non-constant power-law diffusivity has been analysed
using Lie group methods. A classical four-dimensional symmetry group has been
derived which admits the construction of explicit solutions. The analysis of
the associated Lie algebra yields three different classes of equivalent
solutions. Thereby the first and second class lead to the well-known separable
and scale invariant solutions.  To the best of our knowledge, a solution of the
third type has not been reported before. Application of the group transformations 
to the basic solutions yields further two- and three-parameter solutions which
have not been described either.

\section*{Acknowledgements}
I would like to thank R.\ Avramenko for the numerous discussions and comments,
in particular on Lie groups and related topics, which helped to improve the
manuscript a lot.

\bibliography{myrefs}

\begin{thebibliography}{10}
\expandafter\ifx\csname url\endcsname\relax
  \def\url#1{\texttt{#1}}\fi
\expandafter\ifx\csname urlprefix\endcsname\relax\def\urlprefix{URL }\fi
\expandafter\ifx\csname href\endcsname\relax
  \def\href#1#2{#2} \def\path#1{#1}\fi

\bibitem{bluman1974}
G.~W. {Bluman}, J.~D. {Cole}, Similarity methods for differential equations,
  Vol.~13 of Applied mathematical sciences, Springer-Verlag, New York, 1974.

\bibitem{olver1986}
P.~J. {Olver}, Applications of Lie groups to differential equations, Vol. 107
  of Graduate texts in mathematics, Springer-Verlag, New York, 1986.

\bibitem{polyanin2002}
A.~D. {Polyanin}, Handbook of linear partial differential equations for
  engineers and scientists, Chapman \& Hall/CRC, Boca Raton, FL, 2002.

\bibitem{sutton1943}
W.~G.~L. {Sutton}, On the equation of diffusion in a turbulent medium, Proc. R.
  Soc. Lond. A 182~(988) (1943) 48--75.
\newblock \href {https://doi.org/10.1098/rspa.1943.0023}
  {\path{doi:10.1098/rspa.1943.0023}}.

\bibitem{lyndenbell1974}
D.~{Lynden-Bell}, J.~E. {Pringle}, Evolution of viscous disks and origin of
  nebular variables, MNRAS 168~(3) (1974) 603--637.
\newblock \href {https://doi.org/10.1093/mnras/168.3.603}
  {\path{doi:10.1093/mnras/168.3.603}}.

\bibitem{rafikov2016}
R.~R. {Rafikov}, {Generalized Similarity for Accretion/Decretion Disks}, ApJ
  830 (2016) 7.
\newblock \href {http://arxiv.org/abs/1604.07439} {\path{arXiv:1604.07439}},
  \href {https://doi.org/10.3847/0004-637X/830/1/7}
  {\path{doi:10.3847/0004-637X/830/1/7}}.

\bibitem{hydon2000}
P.~E. {Hydon}, {Symmetry methods for differential equations}, Cambridge texts
  in applied mathematics, Cambridge University Press, New York, 2000.

\bibitem{ovsiannikov1982}
L.~V. {Ovsiannikov}, Group analysis of differential equations, Academic Press,
  New York, 1982.

\bibitem{beltrametti1966}
E.~G. {Beltrametti}, A.~{Blasi}, On the number of casimir operators associated
  with any lie group, Physics Letters 20~(1) (1966) 62--64.
\newblock \href {https://doi.org/10.1016/0031-9163(66)91048-1}
  {\path{doi:10.1016/0031-9163(66)91048-1}}.

\bibitem{abramowitz1964}
M.~Abramowitz, I.~A. Stegun, {Handbook of Mathematical Functions}, Vol.~55 of
  Applied Mathematics Series. National Bureau of Standards, US Gov. Print.
  Office, Washington, DC, 1964.

\end{thebibliography}

\end{document}